\documentclass[11pt]{article}
\usepackage{graphicx}
\usepackage{cite}
\usepackage{amssymb}
\usepackage{epstopdf}
\usepackage{hep}
\usepackage{feynmp}
\usepackage{multicol}
\usepackage{simplewick}
\usepackage{appendix}
\usepackage{subfigure}
\usepackage{color}
\usepackage{ulem}
\newcommand{\eins}{\mbox{$1 \hspace{-1.0mm} {\bf l}$}}

\makeatletter
\def\slash#1{{\mathpalette\c@ncel{#1}}} 
\makeatother

\newcommand\beq{\begin{eqnarray}}
\newcommand\eeq{\end{eqnarray}}

\newcommand\la{\langle}
\newcommand\ra{\rangle}

\allowdisplaybreaks

\begin{document}

\title{{\bf Twist-3 fragmentation effects for \boldmath{$A_{LT}$} in light hadron production from proton-proton collisions}}

\author{Y.~Koike$^{1}$, D.~Pitonyak$^{2}$, Y.~Takagi$^{3}$, and S.~Yoshida$^{1}$
 \\[0.3cm]
{\normalsize\it $^1$Department of Physics, Niigata University, Ikarashi, Niigata 950-2181, Japan} \\[0.15cm]
{\normalsize\it $^2$RIKEN BNL Research Center, Brookhaven National Laboratory, Upton, NY 11973, USA} \\[0.15cm]
{\normalsize\it $^3$Graduate School of Science and Technology, Niigata University,
Ikarashi, Niigata 950-2181, Japan} 
}

\date{\today}
\maketitle

\begin{abstract}
\noindent We compute the contribution from the twist-3 fragmentation function for light hadron production in collisions between transversely and longitudinally polarized protons, i.e., $p^\uparrow \vec{p} \to h\,X$, which can cause a double-spin asymmetry (DSA) $A_{LT}$.  This is a na\"{i}ve T-even twist-3 observable that we analyze in collinear factorization using both Feynman gauge and lightcone gauge as well as give a general proof of color gauge invariance.  So far only twist-3 effects in the transversely polarized proton have been studied for $A_{LT}$ in $p^\uparrow \vec{p} \to h\,X$.  However, there are indications that the na\"{i}ve T-odd transverse single-spin asymmetry (SSA) $A_N$ in $p^\uparrow p \to h\,X$ is dominated not by such distribution effects but rather by a fragmentation mechanism.  Therefore, one may expect similarly that the fragmentation contribution is important for $A_{LT}$.  Given possible plans at RHIC to measure this observable, it is timely to provide a calculation of this term. \\


\end{abstract}

%
%
\section{Introduction}
Spin asymmetries in various hard processes have brought a new perspective to high-energy perturbative QCD theory and phenomenology.  Transverse single-spin asymmetries (SSAs) $A_N$, which are na\"{i}ve T-odd observables, were first explored in the mid-1970s in $p^\uparrow p\to \pi\,X$ at Argonne National Lab~\cite{Klem:1976ui} and $p\,Be\to \Lambda^\uparrow X$ at FermiLab~\cite{Bunce:1976yb}.  Both of these measurements led to strikingly large effects that were unexplainable in the na\"{i}ve parton model~\cite{Kane:1978nd}.  Experiments continued at FermiLab in the 1990s for $p^\uparrow p\to \pi\, X$~\cite{Adams:1991rw} and most recently at AGS~\cite{Krueger:1998hz,Allgower:2002qi} and RHIC~\cite{Adams:2003fx,Adler:2005in,Lee:2007zzh,:2008mi,Adamczyk:2012qj,Adamczyk:2012xd,Bland:2013pkt,Adare:2013ekj} for $p^\uparrow p\to \{\pi,\,K,\,jet\}\, X$.  All of their measurements likewise produced substantial transverse SSAs.  On the theoretical side, it was realized in the 1980s by Efremov and Teryaev that if one went beyond the simple parton model and included (collinear twist-3) quark-gluon-quark correlations in the nucleon, then there was the potential to generate these large effects~\cite{Efremov:1981sh}. A systematic approach was then developed
by Qiu and Sterman in the 1990s that presented the collinear twist-3 factorization 
framework~\cite{Qiu:1991pp,Qiu:1991wg,Qiu:1998ia} 
with the expectation that one would be able describe transverse SSAs within this perturbative approach. 
Later a solid foundation was given to this formalism 
in \cite{Eguchi:2006qz,Eguchi:2006mc} that proved the cancelation among gauge-noninvariant
terms and led to an expression for the
twist-3 cross section in terms of the complete set of the twist-3 quark-gluon-quark correlation functions.
Over the last decade, several other analyses, 
including those for the extention to twist-3 fragmentation 
functions\,\cite{Yuan:2009dw,Kang:2010zzb,Metz:2012ct,Kanazawa:2013uia}
and three-gluon correlation functions\,\cite{Beppu:2010qn}, 
furthered the progress of this formalism --- see also \cite{Kouvaris:2006zy,Koike:2006qv,Koike:2007rq,Koike:2009ge,
Kanazawa:2010au,Beppu:2013uda} and references therein.

For many years the main assumption was that these transverse SSAs were due to effects inside the transversely polarized proton, in particular those embodied by the so-called Qiu-Sterman function $T_F$~\cite{Qiu:1991pp,Qiu:1991wg,Qiu:1998ia,Kouvaris:2006zy}.  However, a fit of the QS function to $A_N$ data led to a result that was inconsistent with an extraction of the Sivers function $f_{1T}^\perp$~\cite{Sivers:1989cc} from SIDIS, which has a model-independent relation to $T_F$~\cite{Boer:2003cm}, and became known as the ``sign mismatch'' crisis~\cite{Kang:2011hk}.  An attempt to resolve this issue through more flexible parameterizations of the Sivers function proved unsuccessful~\cite{Kang:2012xf}, and, by looking at $A_N$ data on the target transverse SSA in inclusive DIS~\cite{Airapetian:2009ab,Katich:2013atq}, it was argued in fact that the QS function could not be the main cause of $A_N$~\cite{Metz:2012ui}.  This led to a recent work that examined the impact of fragmentation effects from the outgoing hadron~\cite{Kanazawa:2014dca} based on the analytical calculation in Ref.~\cite{Metz:2012ct}.  It was determined that this fragmentation term could be the dominant source of $A_N$ in $p^\uparrow p\to \pi X$~\cite{Kanazawa:2014dca}. 

In addition to $A_N$, there is another twist-3 observable in proton-proton collisions that can give insight into quark-gluon-quark correlations in the incoming protons and/or outgoing hadron.  This is the longitudinal-transverse double-spin asymmetry (DSA) $A_{LT}$, which, unlike $A_N$, is a na\"{i}ve T-even process.  The classic reaction for which this effect has been analyzed is $A_{LT}$ in inclusive DIS (see~\cite{Posik:2014usi} for recent experimental results on this observable).  This asymmetry has also been studied in the Drell-Yan process involving two incoming polarized hadrons~\cite{Jaffe:1991kp,Tangerman:1994bb,Koike:2008du,Lu:2011th}; in inclusive lepton production from $W$-boson decay in proton-proton scattering~\cite{Metz:2010xs}; for jet production~\cite{Kang:2011jw} and pion production~\cite{Kanazawa:2014tda} in lepton-nucleon collisions; and for direct photon production~\cite{Liang:2012rb}, jet/pion production~\cite{Metz:2012fq}, and $D$-meson production~\cite{Hatta:2013wsa} in proton-proton collisions.  

Of these works on $A_{LT}$, only in Ref.~\cite{Kanazawa:2014tda} for $\vec{\ell}\,p^\uparrow\to \pi\, X$ was the twist-3 fragmentation piece calculated (we will see the structure of that result persists in our computation), whereas the fragmentation term for $p^\uparrow\vec{p}\to \pi\, X$ has never been studied.  Like with $A_N$, there is no reason {\it a priori} that this piece cannot be important or perhaps dominant in the asymmetry.  Given the possible plans by the PHENIX Collaboration at RHIC to measure $A_{LT}$ for pions~\cite{PHENIX:BeamUse}\footnote{We mention that a clear $A_{LT}$ asymmetry has already been seen by the Hall A Collaboration at Jefferson Lab in SIDIS~\cite{Huang:2011bc} and $\vec{\ell}\,n^\uparrow\to \pi\, X$~\cite{Zhao:2015wva}.}, we feel a calculation of the fragmentation term for this final state is needed at this time.  Furthermore, like prior research in the literature~\cite{Yuan:2009dw,Kang:2010zzb,Metz:2012ct,Kanazawa:2013uia,Gamberg:2014eia,Kanazawa:2014tda,Kanazawa:2015jxa}, this work will continue to establish/verify the theoretical techniques used in collinear twist-3 fragmentation calculations.  The remainder of the paper is organized as follows: in Sec.~\ref{s:Def} we introduce the twist-3 fragmentation functions relevant for spin-0 hadron production. Next, in Sec.~\ref{s:Calc} we discuss the calculation of the polarized cross section formula for $A_{LT}$.  Finally, in Sec.~\ref{s:Sum} we summarize our work and give an outlook.  A general proof of the color gauge invariance of our result is given in the Appendix.

%
%
\section{Twist-3 fragmentation functions for spin-0 hadrons \label{s:Def}}

We now define the set of twist-3 fragmentation functions relevant for spin-0 hadron production.
The quark-quark matrix element gives two purely real twist-3 functions, which read
\begin{align}
{1\over N}\sum_{X}&\int{d\lambda\over 2\pi}e^{-i{\lambda\over z}}\la 0|\psi^q_i(0)|P_h;X\ra
\la P_h;X|\bar{\psi}^q_j(\lambda w)|0\ra \nonumber\\
&={M_\chi\over z}(\eins)_{ij}\,\widehat{e}^{\,h/q}_1(z)+{M_\chi\over 2z}(\sigma_{\lambda\alpha}i\gamma_5)_{ij}
\epsilon^{\lambda\alpha wP_h}\,\widehat{e}^{\,h/q}_{\bar{1}}(z)+\cdots,
\end{align}
where $\psi_i$ is a quark field with spinor index $i$, and we use the simplified notation $\epsilon^{\lambda\alpha wP_h}
\equiv \epsilon^{\lambda\alpha\rho\sigma}w_{\rho}P_{h\sigma}$ (with $\epsilon_{0123} = +1$).
The color indices are summed over
and divided by the number of colors $N=3$.  The scale
$M_\chi$ is used to make the functions dimensionless and is on the order of the nucleon mass.\footnote{$M_\chi$ is the scale of nonperturbative chiral-symmetry breaking (CSB), which is said to be on the order of the nucleon mass ($\sim 1\,{\rm GeV}$).  We have chosen this scale instead of the light hadron mass $M_h$ since twist-3 functions representing helicity flip effects are due to nonperturbative CSB whereas $M_h$ for the pseudoscalar mesons represents the explicit CSB due to the quark mass.} The vector $w^{\mu}$ is light-like ($w^2=0$) and satisfies
$P_h\cdot w=1$. We will suppress the gauge-link operators throughout for simplicity.

Next, we introduce the so-called $F$-type quark-gluon-quark twist-3 fragmentation functions. We can define two independent functions as
\begin{align}
{1\over N}\sum_{X}&\int{d\lambda\over 2\pi}\int{d\mu\over 2\pi}
e^{-i{\lambda\over z_1}}e^{-i\mu\left({1\over z}-{1\over z_1}\right)}\la 0|\psi^q_i(0)|P_h;X\ra\la P_h;X|\bar{\psi}^q_j(\lambda w)gF^{\alpha w}(\mu w)|0\ra \nonumber\\
&={M_\chi\over 2z}(\gamma_5\slash{P}_h\gamma_\lambda)_{ij}
\epsilon^{\lambda\alpha wP_h}\widehat{E}^{\,h/q}_{F}(z_1,z)+\cdots, \label{F-type1}\\[0.3cm]
{1\over N}\sum_{X}&\int{d\lambda\over 2\pi}\int{d\mu\over 2\pi}
e^{-i{\lambda\over z_1}}e^{-i\mu\left({1\over z}-{1\over z_1}\right)}\la 0|\bar{\psi}^q_j(\lambda w)\psi^q_i(0)|P_h;X\ra\la P_h;X|gF^{\alpha w}(\mu w)|0\ra \nonumber\\
&={M_\chi\over 2z}(\gamma_5\slash{P}_h\gamma_\lambda)_{ij}
\epsilon^{\lambda\alpha wP_h}\widetilde{E}^{\,h/q}_{F}(z_1,z)+\cdots,
\end{align}
where $F^{\alpha w}(\mu w)$ is the gluon field strength tensor.  We note that both $\widehat{E}_{F}(z_1,z)$ 
and $\widetilde{E}_{F}(z_1,z)$ in general are complex functions.  The correlator $\widehat{E}_{F}(z_1,z)$ has support on $1>z>0$ and $z_1>z$, while $\widetilde{E}_{F}(z_1,z)$ has support on $\frac{1} {z}-\frac{1} {z_1} > 1$, $\frac{1} {z_1} <0$, and $\frac{1} {z} >0$~\cite{Kanazawa:2013uia,Meissner:2008yf}.  

We can consider the so-called $D$-type twist-3 fragmentation
functions $\widehat{E}_{D}(z_1,z)$ by replacing $gF^{\alpha w}(\mu w)$ in (\ref{F-type1}) with a covariant derivative 
$D^{\alpha}({\mu w})=\partial^{\alpha}-igA^{\alpha}(\mu w)$. However, $\widehat{E}_{F}(z_1,z)$ and $\widehat{E}_{D}(z_1,z)$
can be related through the identity,
\beq
\widehat{E}^{\,h/q}_{D}(z_1,z)=P\Bigl({1\over 1/z_1-1/z}\Bigr)\,\widehat{E}^{\,h/q}_{F}(z_1,z)+\delta\Bigl({1\over z_1}-{1\over z}\Bigr)\,\widetilde{e}^{\,h/q}(z)\,,
\label{GIR}
\eeq
where $\widetilde{e}(z)$ is another twist-3 fragmentation function that is pure imaginary and defined as
\begin{align}
{1\over N}\sum_{X}&\int{d\lambda\over 2\pi}e^{-i{\lambda\over z}}\la 0|[\infty w,0]\psi^q_i(0)|P_h;X\ra
\la P_h;X|\bar{\psi}^q_j(\lambda w)[\lambda w,\infty w]|0\ra\overleftarrow{\partial}^{\!\alpha} \nonumber\\
&={M_\chi\over 2z}(\gamma_5\slash{P}_h\gamma_\lambda)_{ij}
\epsilon^{\lambda\alpha wP_h}\,\widetilde{e}^{\,h/q}(z)+\cdots.
\end{align}
Note that we have restored the gauge links $[a,b]$ in order to emphasize that $\overleftarrow{\partial}^{\!\alpha}$ acts on the $\lambda$ dependence in both the quark field $\bar{\psi}^q_j(\lambda w)$ and the gauge link $[\lambda w,\infty w]$.
The $D$-type function $\widehat{E}_{D}(z_1,z)$ has another relation associated with the QCD equation of motion,
\beq
z\int_z^\infty{dz_1\over z_1^{2}}\widehat{E}^{\,h/q}_D(z_1,z)=\widehat{e}^{\,h/q}_1(z)+i\,\widehat{e}^{\,h/q}_{\bar{1}}(z)\,.
\label{EOM}
\eeq
By combining Eqs.~(\ref{GIR}), (\ref{EOM}) we can eliminate the $D$-type function and obtain
\beq
z\int_z^\infty{dz_1\over z_1^{2}}P\Bigl({1\over 1/z_1-1/z}\Bigr)\,\widehat{E}^{\,h/q}_{F}(z_1,z)+z\,\widetilde{e}^{\,h/q}(z)=\widehat{e}^{\,h/q}_1(z)+i\,\widehat{e}^{\,h/q}_{\bar{1}}(z)\,.
\eeq
The real and imaginary parts of the above relation respectively give
\begin{align}
z\int_z^\infty{dz_1\over z_1^{2}}P\Bigl({1\over 1/z_1-1/z}\Bigr)\,\widehat{E}^{\,h/q,\Re}_{F}(z_1,z)&=\widehat{e}^{\,h/q}_1(z)\,, \label{T-even}\\[0.3cm]
z\int_z^\infty{dz_1\over z_1^{2}}P\Bigl({1\over 1/z_1-1/z}\Bigr)\,\widehat{E}^{\,h/q,\Im}_{F}(z_1,z)+z\,\widetilde{e}^{\,h/q,\Im}(z)&=\widehat{e}^{\,h/q}_{\bar{1}}(z)\label{e:Im}\,,
\end{align}
where $\Re$ ($\Im$) indicates the real (imaginary) part of the function.  It was shown that Eq.~(\ref{e:Im}) ensures the gauge invariance of the polarized cross section formula in the case of the transverse SSA in SIDIS~\cite{Kanazawa:2013uia}. 
We will show that Eq.~(\ref{T-even}) plays the same role in the case of the longitudinal-transverse DSA in proton-proton collisions.

%
%
\section{Calculation of the polarized cross section for \boldmath{$A_{LT}$} \label{s:Calc}}

We consider the polarized cross section for the production of a light (spin-0) hadron from the collision between a transversely polarized proton 
and a longitudinally polarized proton,
\beq
p(P,S_{\perp})+p(P',\Lambda)\to h(P_h)+X\,,
\label{pphX}
\eeq
where the momenta and polarizations of the particles are given.  The first non-vanishing contribution to the cross section reads
\begin{align} \label{e: collfac}
d\sigma(P_{h\perp},S_{\perp},\Lambda) &= \,H\otimes f_{a/A(3)}\otimes f_{b/B(2)}\otimes D_{C/c(2)} \nonumber \\
&+ \,H'\otimes f_{a/A(2)}\otimes f_{b/B(3)}\otimes D_{C/c(2)} \nonumber \\
&+ \,H''\otimes f_{a/A(2)}\otimes f_{b/B(2)}\otimes D_{C/c(3)}\,,
\end{align} 
where a sum over partonic channels and parton flavors in each channel is understood.  The labels on the functions indicate the parton/proton (or hadron/parton) species and the twist (e.g., $f_{a/A(3)}$ denotes a twist-$3$ correlator associated with parton $a$ in proton $A$).  These functions are convoluted with hard factors $H,\,H',\,H''$, which are different for each term.  The first term in (\ref{e: collfac}) was already calculated in Ref.~\cite{Metz:2012fq}.  With regards to the second term, for the case of the transverse SSA $A_N$ (where $B$ is now unpolarized), which involves chiral-odd twist-3 unpolarized distributions, the authors of Ref.~\cite{Kanazawa:2000hz} demonstrated this part is negligible because of the smallness of the hard scattering coefficients.  However, in Ref.~\cite{Koike:2008du} the authors found in Drell-Yan for $A_{LT}$ this second term, which involves the chiral-odd 
twist-3 longitudinally polarized distribution, can be as large as the first (chiral-even) term.  Therefore, this second term should be analyzed in the future in order to have a complete result for this observable.
We will now compute the third term, which involves the chiral-odd twist-3 fragmentation functions introduced in Sec.~\ref{s:Def} coupled to the (chiral-odd) transversity function $h_1(x)$ and the helicity distribution $g_1(x)$, both defined in the standard way~\cite{Ralston:1979ys,Jaffe:1991kp,Cortes:1991ja}:
\begin{align}
\int{d\lambda\over 2\pi}e^{i\lambda x}\langle P,S|\bar{\psi}^q_j(0)
\psi^q_i(\lambda n)|P,S\rangle&={1\over 2}\left[(\slash{P})_{ij}\,f_{1}^q(x)
+\Lambda(\gamma^5\slash{P})_{ij}\,g_{1}^q(x) +(\gamma_5\slash{S}_{\perp}\slash{P})_{ij}\,h^q_{1}(x)\right], \\[0.3cm]
\int{d\lambda\over 2\pi}e^{i\lambda x}\la P,S|
F^{\alpha n}(0)F^{\beta n}(\lambda n)|P,S\ra
&=-{x\over 2}\left[g_{\perp}^{\alpha\beta}\,f_1^g(x)
+\Lambda i\epsilon^{\alpha\beta Pn}\,g_1^g(x)\right],
\end{align}
where the unpolarized distribution $f_1(x)$ has been included for completeness.  The vector $n^{\mu}$ is light-like and satisfies
$P\cdot n=1$.

\begin{figure}[t]
\begin{center}
\includegraphics[width=14cm]{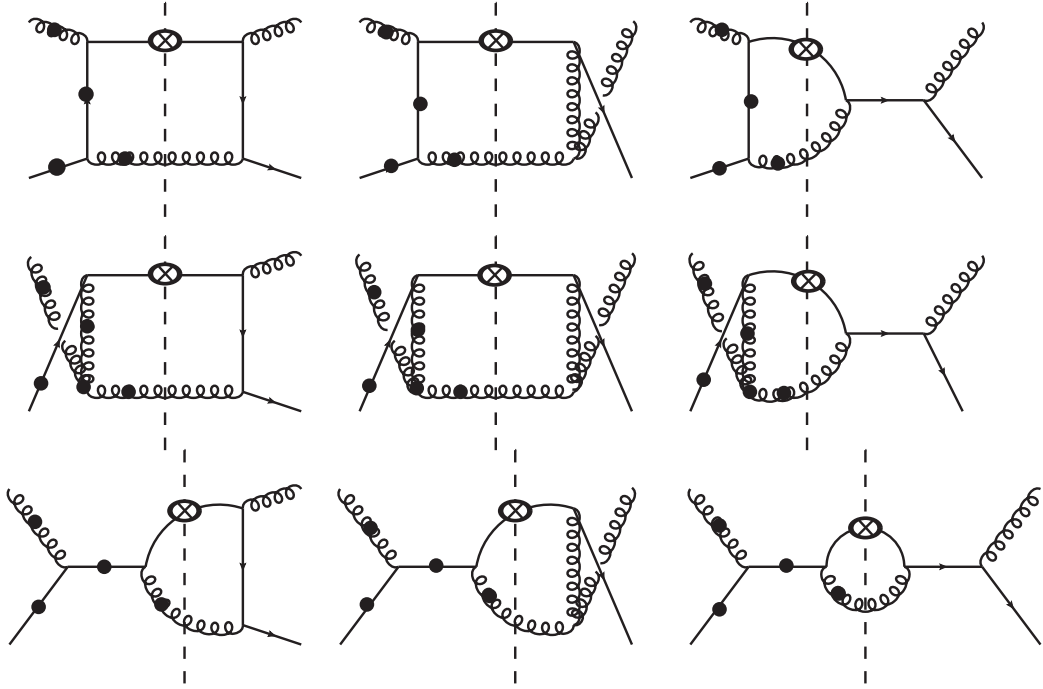}
\vspace{-0.3cm}
\caption[]{Feynman diagrams for the $qg\to qg$ channel. The circled cross indicates the parton that fragments.  The 9 graphs (ignoring the dots) lead to the hard factor for $\widehat{e}_1(z)$ while those with the dots, which represent coherent gluon attachments from the parton line to the fragmentation correlator, give the hard part for $\widehat{E}^{\Re}_F(z_1,z)$.  The Hermitian conjugate (H.c.) graphs are also taken into account.}
 \label{f:qgqg}
\end{center}
\end{figure}

\begin{figure}[t]
\begin{center}
\includegraphics[width=12cm]{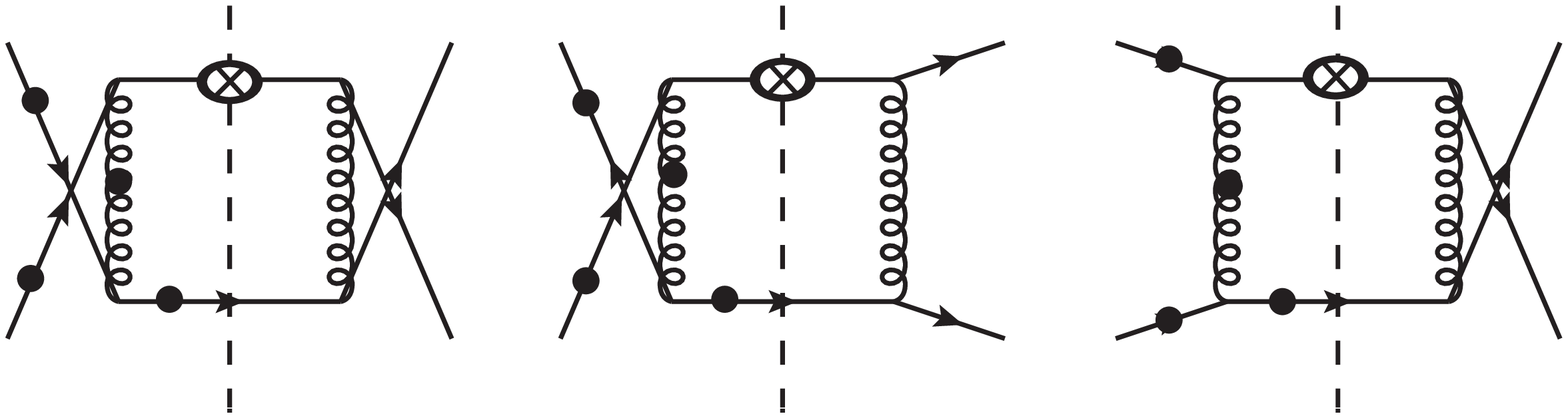}
\vspace{-0.3cm}
\caption[]{Feynman diagrams for  the $qq\to qq$ channel.  The first graph gives the $qq^\prime\to qq^\prime$ channel.}
 \label{f:qqqq}
\end{center}
\end{figure}

\begin{figure}[t]
\begin{center}
\includegraphics[width=14cm]{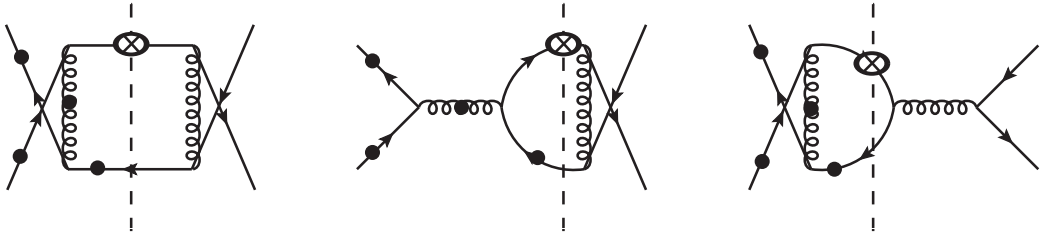}
\vspace{-0.3cm}
\caption[]{Feynman diagrams for the $q\bar{q}\to q\bar{q}$ channel. The first graph gives the $q\bar{q}^{\,\prime}\to q\bar{q}^{\,\prime}$ channel.  Note that the $\bar{q} q\to q\bar{q}$ channel is found by interchanging the two incoming parton lines.}
 \label{f:qqbarqqbar}
\end{center}
\end{figure}

\begin{figure}[t]
\begin{center}
\includegraphics[width=14cm]{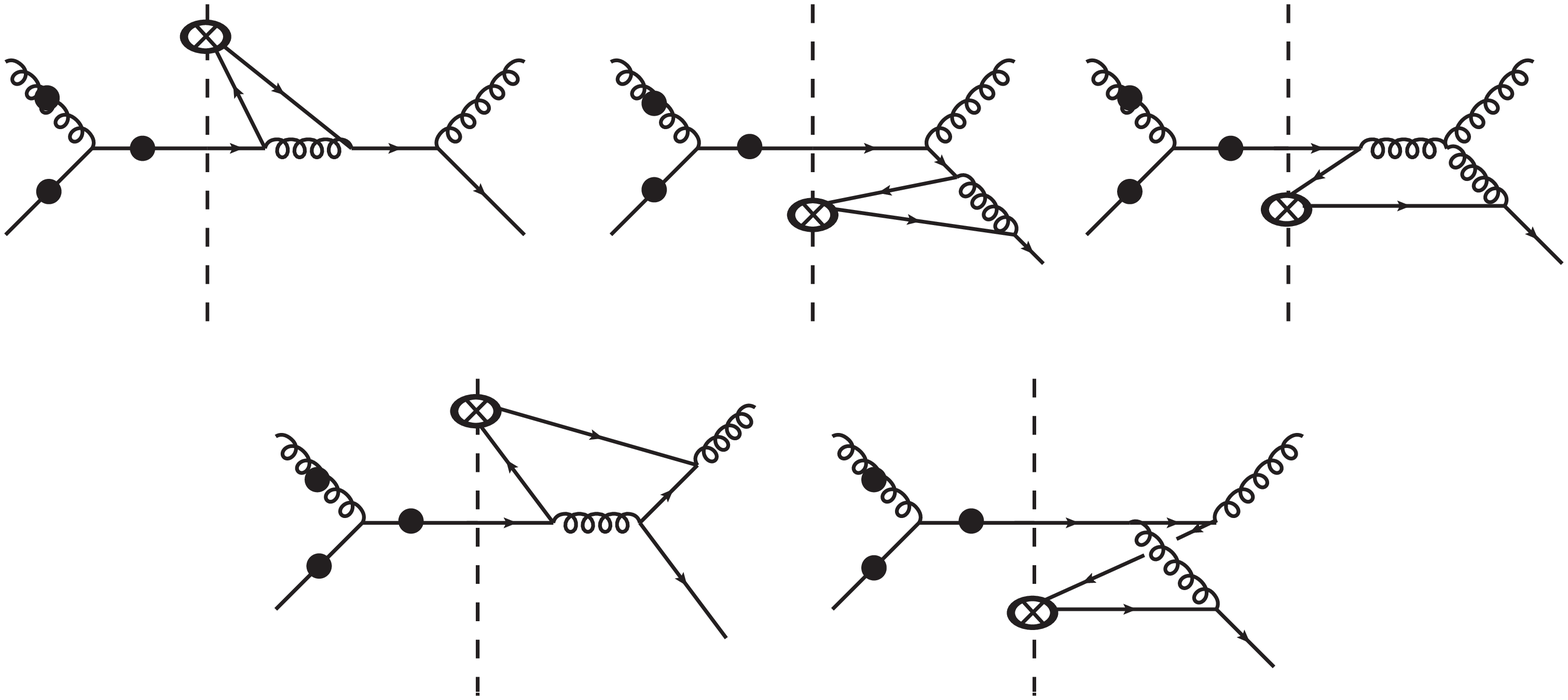}\\
\includegraphics[width=14cm]{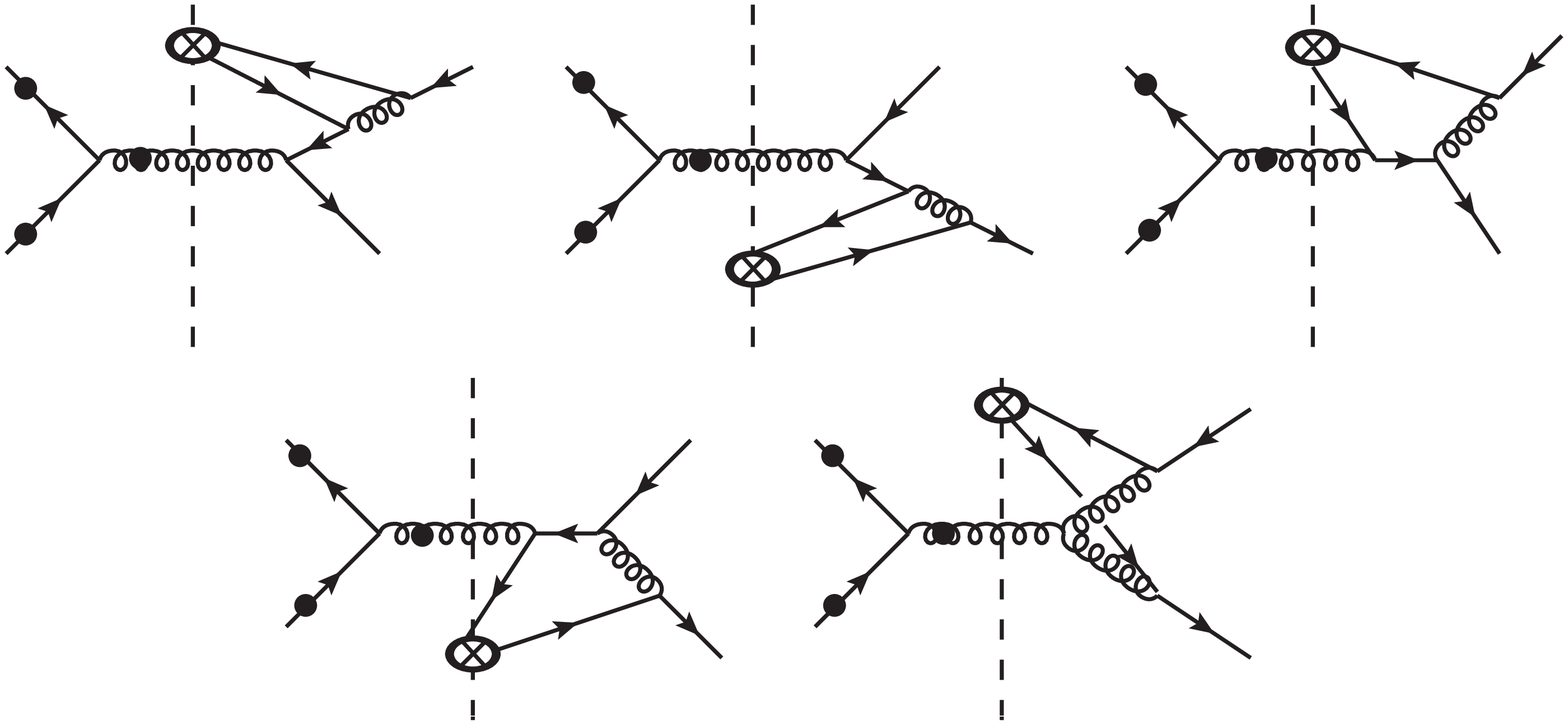}
\vspace{-0.3cm}
\caption[]{Feynman diagrams for $qg$ and $q\bar{q}$ induced channels where both quarks entering the fragmentation are on the same side of the cut.  These lead to the hard factors for $\widetilde{E}_F(z_1,z)$.  Such graphs cancel after one sums all contributions.}
 \label{f:sameside}
\end{center}
\end{figure}

The techniques for calculating the complete fragmentation term in the collinear twist-3 framework have been laid out in Refs.~\cite{Metz:2012ct,Kanazawa:2013uia,Kanazawa:2015jxa}.  In particular, the work in \cite{Metz:2012ct} can be used for deriving the result in lightcone gauge, whereas that in  \cite{Kanazawa:2013uia} can be employed if one chooses Feynman gauge.  For the former, one can make straightforward changes to the calculation in Ref.~\cite{Metz:2012ct} in order to obtain the result for $p^\uparrow \vec{p} \to h\, X$ (since a similar process ($p^\uparrow p \to h\, X$) was computed there).  The channels that one must consider are $qg\to qg$, $qq^\prime\to qq^\prime$, $qq\to qq$, $q\bar{q}\to q\bar{q}$, $q\bar{q}^{\,\prime}\to q\bar{q}^{\,\prime}$, and $\bar{q}q\to q\bar{q}$ (see Figs.~\ref{f:qgqg}--\ref{f:qqbarqqbar}) plus all the antiquark fragmentation channels found through charge conjugating the aforementioned ones.  We note that because of the T-even nature of this observable, the structure of our cross section takes on a different form than $A_N$.  For example, one no longer has contributions from $\widehat{e}_{\bar{1}}(z)$ or $\widetilde{e}(z)$; instead, the quark-quark piece only involves $\widehat{e}_1(z)$.  Likewise, for the quark-gluon-quark part, one receives contributions from $\widehat{E}^{\Re}_F(z_1,z)$ instead of $\widehat{E}^{\Im}_F(z_1,z)$. The hard factors for $\widetilde{E}_F(z_1,z)$ (see Fig.~\ref{f:sameside}) vanish, as was also found in Ref.~\cite{Metz:2012ct} with $A_N$.  Furthermore, unlike $A_N$, it turns out that the quark-gluon-quark hard factors are independent of $z_1$.  Thus, one can use Eq.~(\ref{T-even}) to write this piece in terms of $\widehat{e}_1(z)$, and the entire cross section then only involves this function.  Such a simplification was also noticed in Ref.~\cite{Kanazawa:2014tda} for $\vec{\ell}\,p^\uparrow\to \pi\, X$.  We then find the fragmentation term in the polarized cross section relevant for $A_{LT}$ to be
\begin{equation}
\frac{P_h^0 d\sigma_{LT}^{Frag}} {d^3\vec{P}_h} = -\frac{2\alpha_s^2 M_\chi} {S}\, \Lambda\,P_{h\perp}\cdot S_\perp\,\sum_i\sum_{a,b,c}\int_0^1\!\frac{dz} {z^4}\,\widehat{e}_1^{\,h/c}(z)\,\int_0^1 \!\frac{dx^\prime} {x^\prime}\,g_1^b(x^\prime)\int_0^1\!\frac{dx} {x}\,h_1^a(x) \,\hat{\sigma}_i\,\delta(\hat{s}+\hat{t}+\hat{u})\,. \label{e:ALT_Frag}
\end{equation}
The hard factors $\hat{\sigma}_i$ are given by
\begin{align} 
\hat{\sigma}_{qg\to qg} &= -\frac{3} {2}\left[\frac{\hat{s}^2+\hat{u}^2} {\hat{t}^2\hat{u}} -\frac{1} {N^2}\frac{1} {\hat{u}}\right],\hspace{2cm}
\hat{\sigma}_{qq^\prime\to qq^\prime} = \left(1-\frac{1} {N^2}\right)\!\frac{\hat{s}} {\hat{t}^2}\,,\nonumber \\[0.3cm]
\hat{\sigma}_{qq\to qq} &= \left(1-\frac{1} {N^2}\right)\!\left[\frac{\hat{s}} {\hat{t}^2}+\frac{1} {N}\frac{\hat{s}(\hat{s}-2\hat{t})} {2\hat{t}^2\hat{u}}\right],\hspace{0.81cm}
\hat{\sigma}_{q\bar{q}\to q\bar{q}} = \left(1-\frac{1} {N^2}\right)\!\left[\frac{\hat{s}} {\hat{t}^2}+\frac{1} {N}\frac{\hat{u}-2\hat{t}} {2\hat{t}^2}\right],\label{e:sigma} \\[0.3cm]
\hat{\sigma}_{q\bar{q}^{\prime}\to q\bar{q}^{\prime}} &= \hat{\sigma}_{qq^\prime\to qq^\prime}\,,\hspace{4.6cm}
\hat{\sigma}_{\bar{q}q\to q\bar{q}} =\frac{1} {N}\!\left(1-\frac{1} {N^2}\right)\!\frac{3} {2\hat{u}}\,. \nonumber
\end{align}
The Mandelstam variables for the process are defined as $S = (P+P')^{2}$, $T = (P-P_h)^{2}$, and $U = (P'-P_h)^{2}$, which on the partonic level give $\hat{s} = xx' S$, $\hat{t} = xT/z$, and $\hat{u} = x' U/z$.  For the antiquark fragmentation channels we find (cf.~\cite{Metz:2012ct}) $\hat{\sigma}_{\bar{a}\bar{b}\to\bar{c}\bar{d}} = \hat{\sigma}_{ab\to cd}$, where $\hat{\sigma}_{ab\to cd}$ are given in (\ref{e:sigma}).  We also calculated the polarized cross section in Feynman gauge using the procedure of Ref.~\cite{Kanazawa:2013uia} and found agreement with Eqs.~(\ref{e:ALT_Frag}), (\ref{e:sigma}).  A general proof of color gauge invariance can also be found in the Appendix.

%
%
\section{Summary and outlook \label{s:Sum}}
Transverse SSAs $A_N$ in single-inclusive processes (e.g., $p^\uparrow p\to h\,X$) are twist-3 observables that have been an intense topic of research for close to 40 years.  Large effects have been found that still have an unclear origin.  Recently it has been shown that the fragmentation term in collinear twist-3 factorization could be the main cause of $A_N$ in $p^\uparrow p\to \pi\,X$ at RHIC~\cite{Kanazawa:2014dca}.  In addition, another twist-3 reaction exists that can also lead to information on quark-gluon-quark correlations in protons/hadrons:~the longitudinal-transverse DSA $A_{LT}$ in $p^\uparrow \vec{p}\to h\,X$.  Already two related observables, $A_{LT}$ in SIDIS~\cite{Huang:2011bc} and in $\vec{\ell}\,n^\uparrow\to \pi\,X$~\cite{Zhao:2015wva}, have been measured and nonzero effects have been found.  However, RHIC, with the only source of (independently manipulated) polarized proton beams in the world, has yet to explore $A_{LT}$ in $p^\uparrow \vec{p}\to h\,X$ despite measuring asymmetries for every other combination of proton spins.  Just recently, though, the PHENIX Collaboration has put forth plans to make this measurement~\cite{PHENIX:BeamUse}.  Some work has been done previously on $A_{LT}$ that looked at the twist-3 effects in the polarized proton~\cite{Metz:2012fq}.  Motivated by the potential of twist-3 fragmentation effects to dominate $A_N$~\cite{Kanazawa:2014dca}, and with no reason {\it a priori} that they should be small, we have computed this term now for $A_{LT}$.  We found that the entire result can be written in terms of a single (twist-3) quark-quark fragmentation function $\widehat{e}_1(z)$ and confirmed this using two different gauges.  In the future we plan to perform a detailed numerical study of $A_{LT}$ in $p^\uparrow \vec{p}\to \pi\,X$ in order to further encourage an experiment at RHIC.  

%
%
\section*{Acknowledgments}

This work has been supported by the Grant-in-Aid for
Scientific Research from the Japanese Society of Promotion of Science
under Contract No.~26287040 (Y.K. and S.Y.) and the RIKEN BNL
Research Center (D.P.).

%
%
\section*{Appendix: General proof of color gauge invariance \label{s:App}}
\begin{figure}[t]
\begin{center}
\includegraphics[width=14cm]{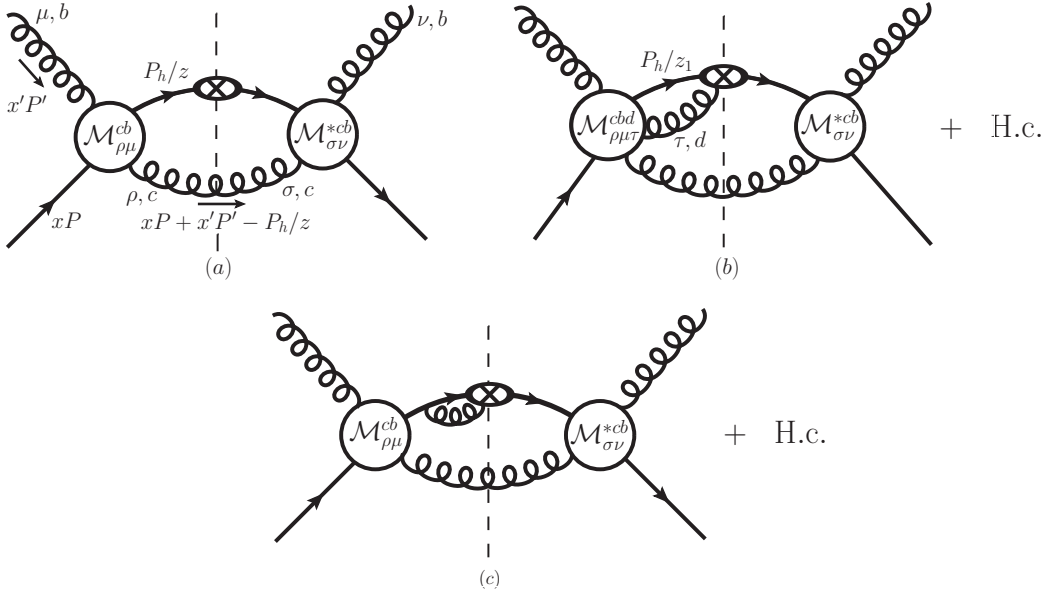}
\vspace{-0.3cm}
\caption[]{Feynman diagrams that enter our general proof of color gauge invariance:~(a) quark-quark and (b), (c) quark-gluon-quark graphs.  Note that our calculation does not include in (b) the subset of diagrams given in (c), but rather these topologies are taken care of by those in (a).  See the text for details.}
 \label{f:GI}
\end{center}
\end{figure}

We want to show that our calculation of the $qg\to qg$ channel satisfies the Ward-Takahashi identity (WTI):
\begin{equation}
(xP+x'P'-P_h/z)^\rho \,d\sigma^{qg\to qg}_{\rho\sigma} = 0\,,
\end{equation}
where $d\sigma^{qg\to qg}_{\rho\sigma}$ is the partonic hard factor with the polarization tensor for the final unobserved gluon $d^{\rho\sigma}\!(xP+x'P'-P_h/z)$ removed.  The momenta and Lorentz indices are shown in Fig.~5(a).  In our computation, we wrote $d\sigma^{qg\to qg}$ as follows (cf.~Fig.~\ref{f:qgqg}):
\begin{equation}
d\sigma^{qg\to qg} = d\sigma^{(a)} + \left(d\sigma^{(b)L} - d\sigma^{(c)L}\right) + \left(d\sigma^{(b)R} - d\sigma^{(c)R}\right)\,,
\end{equation}
where the superscripts correspond to the diagrams in Fig.~\ref{f:GI} ($L$ indicates the explicitly shown graph and $R$ its H.c.).  One sees immediately that 
\begin{equation}
(xP+x'P'-P_h/z)^\rho \,\left\{d\sigma^{(b)L}_{\rho\sigma},\,d\sigma^{(b)R}_{\rho\sigma},\,d\sigma^{(c)R}_{\rho\sigma}\right\}= 0\,,
\end{equation}
since from the WTI for $qg\to qg$ scattering we know
\begin{equation}
(xP+x'P'-P_h/z)^\rho\,\left\{\mathcal{M}_{\rho\mu}^{cb},\,\mathcal{M}_{\rho\mu\tau}^{cbd}\right\} = 0
\end{equation}
if all particles attaching to the respective blob are external (on-shell) lines.  The Dirac projections used for the correlators in $d\sigma^{(b)L}$, $d\sigma^{(b)R}$, $d\sigma^{(c)R}$ allow these connecting particles to meet this criteria.\footnote{The diagram in Fig.~5(a) does not satisfy this requirement because the fragmentation correlator is projected out with $\eins$.}  Therefore, we now have
\begin{align}
(xP+x'P'-P_h/z)^\rho \,d\sigma^{qg\to qg}_{\rho\sigma} &=(xP+x'P'-P_h/z)^\rho\left( d\sigma^{(a)}_{\rho\sigma} - d\sigma_{\rho\sigma}^{(c)L}\right) \nonumber\\[0.3cm]
&\sim (xP+x'P'-P_h/z)^\rho\,\epsilon^{\mu\nu Pn}\nonumber\\
&\hspace{1cm}\times\,\Bigg\{\!-\frac{i} {2}\,\epsilon^{\tau\tau' P_h w}\,{\rm Tr}\!\left[\slashed{S}_\perp\slashed{P}\gamma_5\mathcal{M}_{\sigma\nu}^{*cb}\gamma_{\tau'}\!\left(\slashed{P}_h/z\right)\!\gamma_5\gamma_\tau\frac{\slashed{P}_h/z} {(P_h/z)^2}\mathcal{M}_{\rho\mu}^{cb}\right] \nonumber\\
&\hspace{2cm}+\, {\rm Tr}\!\left[\slashed{S}_\perp\slashed{P}\gamma_5\mathcal{M}_{\sigma\nu}^{*cb}\mathcal{M}_{\rho\mu}^{cb}\right]\!\Bigg\}.
\end{align}
One can show that
\begin{align}
(xP+x'P'-P_h/z)^\rho\,&\epsilon^{\mu\nu Pn}\,\mathcal{M}_{\rho\mu}^{cb}\,\slashed{P}\nonumber\\
&\sim \epsilon^{\mu\nu Pn}\left(\slashed{P}_h/z\right)\!\left[\frac{i(x\slashed{P}+x'\slashed{P}')} {\hat{s}}\,\gamma_\mu\,T^c\,T^b+\frac{P_{h\mu}/z}{\hat{t}}\,f^{bdc}\,T^d\right]\!\slashed{P}\nonumber\\[0.3cm]
&\equiv \left(\slashed{P}_h/z\right)\mathcal{M}^{'\nu cb}\slashed{P}\,,
\end{align}
where $f^{abc}$ are the structure constants and $T^a$ the generators of $SU(3)$.  This leads to
\begin{align}
(xP+x'P'-P_h/z)^\rho \,d\sigma^{qg\to qg}_{\rho\sigma} &\sim -\frac{i} {2}\,\epsilon^{\tau\tau' P_h w}\,{\rm Tr}\!\left[\slashed{S}_\perp\slashed{P}\gamma_5\mathcal{M}_{\sigma\nu}^{*cb}\gamma_{\tau'}\!\left(\slashed{P}_h/z\right)\!\gamma_5\gamma_\tau\mathcal{M}^{'\nu cb}\right] \nonumber\\[0.1cm]
&\hspace{0.5cm}+\, {\rm Tr}\!\left[\slashed{S}_\perp\slashed{P}\gamma_5\mathcal{M}_{\sigma\nu}^{*cb}\!\left(\slashed{P}_h/z\right)\!\mathcal{M}^{'\nu cb}\right].
\end{align}
We can simplify this expression further through use of the identities
\beq
\gamma_5\sigma_{\mu\nu}=-{i\over 2}\epsilon_{\mu\nu}^{\ \ \alpha\beta}
\sigma_{\alpha\beta}\,,\ \ \ \ 
\epsilon^{\mu\nu\rho\sigma}\epsilon_{\mu\nu}^{\ \ \alpha\beta}
=-2(g^{\rho\alpha}g^{\sigma\beta}-g^{\rho\beta}g^{\sigma\alpha})
\eeq
and obtain
\begin{align}
(xP+x'P'-P_h/z)^\rho \,d\sigma^{qg\to qg}_{\rho\sigma} &\sim \frac{1} {2}\,{\rm Tr}\!\left[\slashed{S}_\perp\slashed{P}\gamma_5\mathcal{M}_{\sigma\nu}^{*cb}\!\left(\slashed{P}_h/z\right)\!\!\left(\slashed{P}_h\slashed{w}-\slashed{w}\slashed{P}_h\right)\!\mathcal{M}^{'\nu cb}\right] \nonumber\\
&\hspace{0.5cm}+\, {\rm Tr}\!\left[\slashed{S}_\perp\slashed{P}\gamma_5\mathcal{M}_{\sigma\nu}^{*cb}\!\left(\slashed{P}_h/z\right)\!\mathcal{M}^{'\nu cb}\right]\nonumber\\
&= 0\,.
\end{align}
Thus, the $qg\to qg$ channel satisfies the WTI.


\begin{thebibliography}{99}

\bibitem{Klem:1976ui} 
  R.~D.~Klem, J.~E.~Bowers, H.~W.~Courant, H.~Kagan, M.~L.~Marshak, E.~A.~Peterson, K.~Ruddick and W.~H.~Dragoset {\it et al.},
  Phys.\ Rev.\ Lett.\  {\bf 36}, 929 (1976).
  
\bibitem{Bunce:1976yb} 
  G.~Bunce {\it et al.},
  Phys.\ Rev.\ Lett.\  {\bf 36}, 1113 (1976).
  
\bibitem{Kane:1978nd} 
  G.~L.~Kane, J.~Pumplin and W.~Repko,
  Phys.\ Rev.\ Lett.\  {\bf 41}, 1689 (1978).
  
\bibitem{Adams:1991rw} 
  D.~L.~Adams {\it et al.} [E581 and E704 Collaborations],
  Phys.\ Lett.\ B {\bf 261}, 201 (1991);
  D.~L.~Adams {\it et al.} [E704 Collaboration],
  Phys.\ Lett.\ B {\bf 264}, 462 (1991).

\bibitem{Krueger:1998hz} 
  K.~Krueger {\it et al.},
  Phys.\ Lett.\ B {\bf 459}, 412 (1999).
  
\bibitem{Allgower:2002qi} 
  C.~E.~Allgower, K.~W.~Krueger, T.~E.~Kasprzyk, H.~M.~Spinka, D.~G.~Underwood, A.~Yokosawa, G.~Bunce and H.~Huang {\it et al.},
  Phys.\ Rev.\ D {\bf 65}, 092008 (2002).
  
\bibitem{Adams:2003fx}
  J.~Adams {\it et al.} [STAR Collaboration],
  Phys.\ Rev.\ Lett.\  {\bf 92}, 171801 (2004)
  [hep-ex/0310058];
  B.~I.~Abelev {\it et al.} [STAR Collaboration],
  Phys.\ Rev.\ Lett.\  {\bf 101}, 222001 (2008)
  [arXiv:0801.2990 [hep-ex]].

\bibitem{Adler:2005in} 
  S.~S.~Adler {\it et al.}  [PHENIX Collaboration],
  Phys.\ Rev.\ Lett.\  {\bf 95}, 202001 (2005)
  [hep-ex/0507073].
  
\bibitem{Lee:2007zzh} 
  J.~H.~Lee {\it et al.}  [BRAHMS Collaboration],
  AIP Conf.\ Proc.\  {\bf 915}, 533 (2007).

\bibitem{:2008mi} 
  I.~Arsene {\it et al.}  [BRAHMS Collaboration],
  Phys.\ Rev.\ Lett.\  {\bf 101}, 042001 (2008)
  [arXiv:0801.1078 [nucl-ex]].
  
\bibitem{Adamczyk:2012qj} 
  L.~Adamczyk {\it et al.}  [STAR Collaboration],
  Phys.\ Rev.\ D {\bf 86}, 032006 (2012)
  [arXiv:1205.2735 [nucl-ex]].
  
\bibitem{Adamczyk:2012xd} 
  L.~Adamczyk {\it et al.}  [STAR Collaboration],
  Phys.\ Rev.\ D {\bf 86}, 051101 (2012)
  [arXiv:1205.6826 [nucl-ex]].
  
\bibitem{Bland:2013pkt} 
  L.~C.~Bland {\it et al.}  [AnDY Collaboration],
  arXiv:1304.1454 [hep-ex].
  
\bibitem{Adare:2013ekj} 
  A.~Adare {\it et al.}  [PHENIX Collaboration],
  Phys.\ Rev.\ D {\bf 90}, 012006 (2014)
  [arXiv:1312.1995 [hep-ex]];
  Phys.\ Rev.\ D {\bf 90}, 072008 (2014)
  [arXiv:1406.3541 [hep-ex]].
  
\bibitem{Efremov:1981sh}
  A.~V.~Efremov and O.~V.~Teryaev,
  Sov.\ J.\ Nucl.\ Phys.\  {\bf 36}, 140 (1982)
  [Yad.\ Fiz.\  {\bf 36}, 242 (1982)];
  Phys.\ Lett.\ B {\bf 150}, 383 (1985).
  
\bibitem{Qiu:1991pp} 
  J.-w.~Qiu and G.~F.~Sterman,
  Phys.\ Rev.\ Lett.\  {\bf 67}, 2264 (1991).

\bibitem{Qiu:1991wg} 
  J.-w.~Qiu and G.~F.~Sterman,
  Nucl.\ Phys.\ B {\bf 378}, 52 (1992).

\bibitem{Qiu:1998ia} 
  J.-w.~Qiu and G.~F.~Sterman,
  Phys.\ Rev.\ D {\bf 59}, 014004 (1999)
 [hep-ph/9806356].
  
\bibitem{Eguchi:2006qz}
  H.~Eguchi, Y.~Koike and K.~Tanaka,
  Nucl.\ Phys.\  B {\bf 752}, (2006)
  [arXiv:hep-ph/0604003].

\bibitem{Eguchi:2006mc}
  H.~Eguchi, Y.~Koike and K.~Tanaka,
  Nucl.\ Phys.\  B {\bf 763}, 198 (2007)
  [arXiv:hep-ph/0610314].

\bibitem{Yuan:2009dw} 
  F.~Yuan and J.~Zhou,
  Phys.\ Rev.\ Lett.\  {\bf 103}, 052001 (2009)
  [arXiv:0903.4680 [hep-ph]].
  
\bibitem{Kang:2010zzb} 
  Z.~B.~Kang, F.~Yuan and J.~Zhou,
  Phys.\ Lett.\ B {\bf 691}, 243 (2010)
  [arXiv:1002.0399 [hep-ph]].

\bibitem{Metz:2012ct} 
  A.~Metz and D.~Pitonyak,
  Phys.\ Lett.\ B {\bf 723}, 365 (2013)
  [arXiv:1212.5037 [hep-ph]].
  
\bibitem{Kanazawa:2013uia} 
  K.~Kanazawa and Y.~Koike,
  Phys.\ Rev.\ D {\bf 88}, 074022 (2013)
  [arXiv:1309.1215 [hep-ph]].
 
\bibitem{Beppu:2010qn} 
  H.~Beppu, Y.~Koike, K.~Tanaka and S.~Yoshida,
  Phys.\ Rev.\ D {\bf 82}, 054005 (2010)
  [arXiv:1007.2034 [hep-ph]].  
  

\bibitem{Kouvaris:2006zy}
  C.~Kouvaris, J.~W.~Qiu, W.~Vogelsang and F.~Yuan,
  Phys.\ Rev.\  D {\bf 74}, 114013 (2006)
 [arXiv:hep-ph/0609238].
  
\bibitem{Koike:2006qv} 
  Y.~Koike and K.~Tanaka,
  Phys.\ Lett.\ B {\bf 646}, 232 (2007)
  [Erratum-ibid.\ B {\bf 668}, 458 (2008)]
  [hep-ph/0612117].
  
\bibitem{Koike:2007rq} 
  Y.~Koike and K.~Tanaka,
  Phys.\ Rev.\ D {\bf 76}, 011502 (2007)
  [hep-ph/0703169].

\bibitem{Koike:2009ge} 
  Y.~Koike and T.~Tomita,
  Phys.\ Lett.\ B {\bf 675}, 181 (2009)
 [arXiv:0903.1923 [hep-ph]].

\bibitem{Kanazawa:2010au} 
  K.~Kanazawa and Y.~Koike,
  Phys.\ Rev.\ D {\bf 82}, 034009 (2010)
  [arXiv:1005.1468 [hep-ph]];
  Phys.\ Rev.\ D {\bf 83}, 114024 (2011)
  [arXiv:1104.0117 [hep-ph]].
   
\bibitem{Beppu:2013uda} 
  H.~Beppu, K.~Kanazawa, Y.~Koike and S.~Yoshida,
  Phys.\ Rev.\ D {\bf 89}, 034029 (2014)
  [arXiv:1312.6862 [hep-ph]].
  
\bibitem{Sivers:1989cc} 
  D.~W.~Sivers,
  Phys.\ Rev.\ D {\bf 41}, 83 (1990);
  Phys.\ Rev.\ D {\bf 43}, 261 (1991).
  
\bibitem{Boer:2003cm} 
  D.~Boer, P.~J.~Mulders and F.~Pijlman,
  Nucl.\ Phys.\ B {\bf 667}, 201 (2003).
  
\bibitem{Kang:2011hk} 
  Z.~B.~Kang, J.~W.~Qiu, W.~Vogelsang and F.~Yuan,
  Phys.\ Rev.\ D {\bf 83}, 094001 (2011)
  [arXiv:1103.1591 [hep-ph]].
  
\bibitem{Kang:2012xf} 
  Z.-B.~Kang and A.~Prokudin,
  Phys.\ Rev.\ D {\bf 85}, 074008 (2012)
  [arXiv:1201.5427 [hep-ph]].
  
\bibitem{Airapetian:2009ab} 
  A.~Airapetian {\it et al.}  [HERMES Collaboration],
  Phys.\ Lett.\ B {\bf 682}, 351 (2010)
  [arXiv:0907.5369 [hep-ex]].
  
\bibitem{Katich:2013atq} 
  J.~Katich {\it et al.}
  Phys.\ Rev.\ Lett.\  {\bf 113}, 022502 (2014)
  [arXiv:1311.0197 [nucl-ex]].
  
\bibitem{Metz:2012ui} 
  A.~Metz, D.~Pitonyak, A.~Sch\"{a}fer, M.~Schlegel, W.~Vogelsang and J.~Zhou,
  Phys.\ Rev.\ D {\bf 86}, 094039 (2012)
  [arXiv:1209.3138 [hep-ph]].
  
\bibitem{Kanazawa:2014dca} 
  K.~Kanazawa, Y.~Koike, A.~Metz and D.~Pitonyak,
  Phys.\ Rev.\ D {\bf 89}, 111501(R) (2014)
  [arXiv:1404.1033 [hep-ph]].
  
\bibitem{Posik:2014usi} 
  M.~Posik {\it et al.}  [Jefferson Lab Hall A Collaboration],
  Phys.\ Rev.\ Lett.\  {\bf 113}, 022002 (2014)
  [arXiv:1404.4003 [nucl-ex]].
  
\bibitem{Jaffe:1991kp}
  R.~L.~Jaffe and X.~D.~Ji,
  Phys.\ Rev.\ Lett.\  {\bf 67}, 552 (1991);
  Nucl.\ Phys.\  B {\bf 375}, 527 (1992).
  
\bibitem{Tangerman:1994bb} 
  R.~D.~Tangerman and P.~J.~Mulders,
  hep-ph/9408305.
  
\bibitem{Koike:2008du} 
  Y.~Koike, K.~Tanaka and S.~Yoshida,
  Phys.\ Lett.\ B {\bf 668}, 286 (2008)
  [arXiv:0805.2289 [hep-ph]].
  
\bibitem{Lu:2011th} 
  Z.~Lu and I.~Schmidt,
  Phys.\ Rev.\ D {\bf 84}, 114004 (2011)
  [arXiv:1109.3232 [hep-ph]].
  
\bibitem{Metz:2010xs} 
  A.~Metz and J.~Zhou,
  Phys.\ Lett.\ B {\bf 700}, 11 (2011)
  [arXiv:1006.3097 [hep-ph]].
  
\bibitem{Kang:2011jw} 
  Z.~B.~Kang, A.~Metz, J.~W.~Qiu and J.~Zhou,
  Phys.\ Rev.\ D {\bf 84}, 034046 (2011)
  [arXiv:1106.3514 [hep-ph]].
  
\bibitem{Kanazawa:2014tda} 
  K.~Kanazawa, A.~Metz, D.~Pitonyak and M.~Schlegel,
  Phys.\ Lett.\ B {\bf 742}, 340 (2015)
  [arXiv:1411.6459 [hep-ph]].
  
\bibitem{Liang:2012rb} 
  Z.~T.~Liang, A.~Metz, D.~Pitonyak, A.~Sch\"{a}fer, Y.~K.~Song and J.~Zhou,
  Phys.\ Lett.\ B {\bf 712}, 235 (2012)
  [arXiv:1203.3956 [hep-ph]].
  
\bibitem{Metz:2012fq} 
  A.~Metz, D.~Pitonyak, A.~Sch\"{a}fer and J.~Zhou,
  Phys.\ Rev.\ D {\bf 86}, 114020 (2012)
  [arXiv:1210.6555 [hep-ph]].
  
\bibitem{Hatta:2013wsa} 
  Y.~Hatta, K.~Kanazawa and S.~Yoshida,
  Phys.\ Rev.\ D {\bf 88}, 014037 (2013)
  [arXiv:1305.7001 [hep-ph]].
  
\bibitem{PHENIX:BeamUse} 
  PHENIX Beam Use Proposal: Run-16 and Run-17, May 2015.
  
\bibitem{Huang:2011bc} 
  J.~Huang {\it et al.} [Jefferson Lab Hall A Collaboration],
  Phys.\ Rev.\ Lett.\  {\bf 108}, 052001 (2012)
  [arXiv:1108.0489 [nucl-ex]].
  
\bibitem{Zhao:2015wva} 
  Y.~X.~Zhao {\it et al.} [Jefferson Lab Hall A Collaboration],
  Phys.\ Rev.\ C {\bf 92}, 015207 (2015)
  [arXiv:1502.01394 [nucl-ex]].
      
\bibitem{Gamberg:2014eia} 
  L.~Gamberg, Z.~B.~Kang, A.~Metz, D.~Pitonyak and A.~Prokudin,
  Phys.\ Rev.\ D {\bf 90}, 074012 (2014)
  [arXiv:1407.5078 [hep-ph]].
  
\bibitem{Kanazawa:2015jxa} 
  K.~Kanazawa, A.~Metz, D.~Pitonyak and M.~Schlegel,
  Phys.\ Lett.\ B {\bf 744}, 385 (2015)
  [arXiv:1503.02003 [hep-ph]].
  
\bibitem{Meissner:2008yf} 
  S.~Meissner and A.~Metz,
  Phys.\ Rev.\ Lett.\  {\bf 102}, 172003 (2009)
  [arXiv:0812.3783 [hep-ph]].
  
\bibitem{Kanazawa:2000hz} 
  Y.~Kanazawa and Y.~Koike,
  Phys.\ Lett.\ B {\bf 478}, 121 (2000)
  [hep-ph/0001021];
  Phys.\ Lett.\ B {\bf 490}, 99 (2000)
  [hep-ph/0007272].
  
\bibitem{Ralston:1979ys} 
  J.~P.~Ralston and D.~E.~Soper,
  Nucl.\ Phys.\ B {\bf 152}, 109 (1979).
  
\bibitem{Cortes:1991ja} 
  J.~L.~Cortes, B.~Pire and J.~P.~Ralston,
  Z.\ Phys.\ C {\bf 55}, 409 (1992).
  
  \end{thebibliography}
\end{document}